\documentclass[useAMS,usenatbib]{mn2e}
\usepackage{epsfig}
\usepackage{amsmath}

\newcommand{\be}{\begin{equation}}
\newcommand{\beq}{\begin{equation}}
\newcommand{\ba}{\begin{eqnarray}}
\newcommand{\ee}{\end{equation}}
\newcommand{\eeq}{\end{equation}}
\newcommand{\ea}{\end{eqnarray}}

\newcommand{\apj}{ApJ}

\def\lsim{~\rlap{$<$}{\lower 1.0ex\hbox{$\sim$}}}

\def\gsim{~\rlap{$>$}{\lower 1.0ex\hbox{$\sim$}}}

\title[Smooth Boundaries to Cosmological HII Regions from Galaxy
Clustering]{Smooth Boundaries to Cosmological HII regions from Galaxy
Clustering}

\author[Wyithe \& Loeb]{J. Stuart B. Wyithe$^1$ and Abraham Loeb$^2$\\$^1$ School of Physics, University of Melbourne, Parkville, Victoria, Australia\\$^2$ Harvard-Smithsonian Center for Astrophysics, 60 Garden
St., Cambridge, MA 02138\\Email: swyithe@physics.unimelb.edu.au, loeb@cfa.harvard.edu}

\begin{document}

%\date{\today}
%\pagerange{\pageref{firstpage}--\pageref{lastpage}} \pubyear{2006}

\maketitle

\label{firstpage}
\begin{abstract}

The HII regions around quasars and galaxies at redshifts beyond the epoch
of reionisation will provide prime targets for upcoming 21cm campaigns
using a new generation of low-frequency radio observatories. Here we show
that the boundaries of these HII regions will not be sharp. Rather, the
clustering of sources near massive galaxies results in a neutral fraction
that rises gradually towards large radii from an interior value near zero.
A neutral fraction corresponding to the global background value is
typically reached at a distance of 2--5 times the radius of the HII region
around the central massive galaxy.

\end{abstract}


\section{Introduction}

The identification of a Gunn-Peterson trough and ionised bubbles in the
spectra of quasars at redshifts of $z\sim 6.3$--$6.4$ (Fan et al.~2004)
hints that the reionisation of cosmic hydrogen was completed only around
that time, about a billion years after the big bang (White et al. 2003;
Wyithe \& Loeb~2004).  In the near future, the best probe of the epoch
beyond reionisation will be provided by a new generation of low-frequency
radio arrays that will image the redshifted 21cm emission from the neutral
hydrogen in the intergalactic medium (IGM)\footnote{See
http://www.haystack.mit.edu/ast/arrays/mwa/site/index.html;
http://www.lofar.org/; http://arxiv.org/abs/astro-ph/0502029}. The first
target for these observations will be provided by the cosmological HII
regions generated by massive high redshift galaxies and quasars. These
cosmological HII regions are generally modeled with sharp boundaries within
an IGM having a global neutral hydrogen fraction. However clustering of
small galaxies around massive systems could lead to enhanced ionisation in
the IGM immediately surrounding HII regions. In this {\it Letter} we
estimate this effect. Our approach is to use our earlier model for HII
regions around clustered sources (Wyithe \& Loeb~2005; hereafter Paper I).
Throughout the paper we adopt the set of cosmological parameters determined
by {\it WMAP} (Spergel et al. 2006) for a flat $\Lambda$CDM universe.

\section{The Size of Ionised Regions Around Isolated Galaxies}

We begin by summarising the features of our model for HII regions
around clustered galaxies. Only a brief description is given here, and we
direct the reader to Paper I for further details. There are two
contributions to the ionising luminosity of a galaxy, which we describe in
turn: massive stars and a central supermassive black-hole.

In a neutral IGM the characteristic physical size of a spherical HII region generated by
stars in a galaxy with a dark matter halo of mass $M_{\rm halo}$ is given
by (Loeb et al.~2005)
\begin{eqnarray}
\label{Dstar}
\nonumber
\nonumber D_{\star} &=& 0.75\left(\frac{M_{\rm
halo}}{10^{10}M_\odot}\right)^{1/3}\left(\frac{f_{\star}f_{\rm esc}/N_{\rm
reion}}{0.003}\right)^{1/3}\\
&\times&\left(\frac{N_{\star}}{4000}\right)^{1/3}
\left(\frac{1+z}{7.5}\right)^{-1}\mbox{Mpc},
\end{eqnarray}
where $f_{\star}$ is the fraction of baryons in the halo that are turned
into stars, and $f_{\rm esc}$ is the fraction of ionising photons that
escape from the galaxy into the surrounding IGM.  The number of ionising photons per baryon
incorporated into Pop-II stars is $N_{\star}\sim4000$ (Bromm, Kudritzki \&
Loeb~2001), and $N_{\rm reion}$ denotes the number of photons required per
baryon for the reionisation of the HII region.  The value of $N_{\rm
reion}$ depends on the number of recombinations and hence on the clumpiness
of the IGM.  The halo mass $M_{\rm halo}$ is a function of its virial
velocity dispersion $\sigma$, which we set equal to $v_{\rm c}/\sqrt{2}$,
where $v_c$ is the virial circular velocity [see Eqs.~(22)--(25) in Barkana
\& Loeb~(2001)].

Similarly, the physical size of the HII region generated by an accreting
supermassive black-hole of mass $M_{\rm bh}$ is (White et al., 2003)
\begin{eqnarray}
\label{Dquasar}
\nonumber
D_{\rm q} &=& 0.43\left(\frac{M_{\rm bh}}{10^5M_\odot}\right)^{1/3} \left(\frac{{\dot N}_{\rm
\gamma,5}/N_{\rm reion}}{3\times10^{53}\mbox{s}^{-1}}\right)^{1/3}\\
&\times&
\left(\frac{f_{\rm dyn}}{0.035}\right)^{1/3}
\left(\frac{1+z}{7.5}\right)^{-1}\mbox{Mpc},
\end{eqnarray}
where ${\dot N}_{\rm \gamma,5}$ is the production rate of ionising photons
during the luminous quasar phase for a black hole of mass $M_{\rm
bh}=10^5M_\odot$, and $f_{\rm dyn}=0.035$ is the fraction of the halo
dynamical time during which the quasar shines at its peak luminosity. The
black hole mass is related to the halo velocity dispersion through $M_{\rm
bh} = 10^5 \left({\sigma}/{54~{\rm km~s^{-1}}}\right)^5M_\odot$ (Wyithe \&
Loeb~2003).

\section{The Structure of HII Regions Around Clustered Galaxies} 
\label{RHII}

\begin{figure*}
\includegraphics[width=14cm]{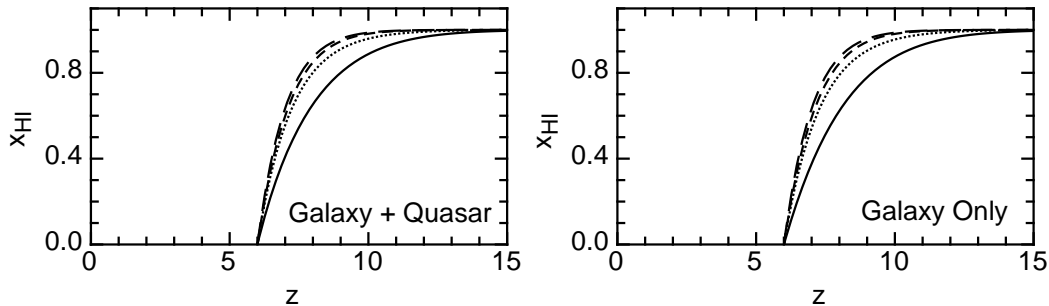} \caption{ The mean neutral
  fraction of the IGM, $x_{\rm HI}$, as a function of redshift $z$ for the
  four models of star formation {\bf A}--{\bf D} (see text for details).}
\label{fig1}
\end{figure*}

Prior to reionisation, isolated galaxies generate their own Str\"omgren
(HII) spheres with characteristic sizes given by equations~(\ref{Dstar})
and (\ref{Dquasar}). However, massive galaxies tend to form in high density
environments within which many smaller galaxies are clustered and contribute to the same HII
region. 

Fixing the center of coordinates to be at the center of largest local galaxy, we may use equations~(\ref{Dstar}) and (\ref{Dquasar}) to find 
the physical size $R_{\rm s}$ of the HII region surrounding a galaxy of
mass $M_{\rm halo}$. We obtain
\begin{equation}
\label{RS}
R_{\rm s}^3 = D_{\rm halo}^3+\int_{R_{\rm vir}}^{R_{\rm s}}4\pi R^2
dR \int_{M_{\rm min}}^{M_{\rm halo}}dM \frac{D_{\rm M}^3}{\Delta_{\rm
R}^2}\frac{dn\left(R,M_{\rm halo}\right)}{dM},
\end{equation}  
where the radii of the HII regions around the central halo of mass $M_{\rm
halo}$ and a galaxy of mass $M$, respectively, are $D_{\rm halo}^3=D_{\rm
halo,q}^3+D_{\rm halo,\star}^3$ and $D_{\rm M}^3=D_{\rm M,q}^3+D_{\rm
M,\star}^3$. Equation~(\ref{RS}) determines $R_{\rm s}$ self-consistently
with the mean neutral fraction into which the HII region expands, and
accounts for the higher IGM density due to infall around a massive halo
through the nonlinear density contrast $\Delta_{\rm R}(R,M_{\rm
halo})$. The number of galaxies surrounding the central halo is found from
the Press-Schechter~(1974) mass function of halos $dn(R,M_{\rm halo})/dM$
with the modification of Sheth \& Torman~(2002), expressed in units
of (physical Mpc)$^{-3}M_\odot^{-1}$.

In the vicinity of a massive halo $dn/dM$ must be modified relative to the
background universe (see Paper I). The critical value of linear overdensity
at which the actual collapse of a spherical shell occurs ($\delta_{\rm
crit}=1.69$ for $\Omega_m=1$) depends weakly on the background matter density
(Bryan \& Norman~1998). The dominant contribution to the modification of
the halo mass function originates from the accelerated formation of halos
in overdense regions. 

In the background universe a spherical top-hat
collapses at redshift $z$ if its overdensity extrapolated (linearly) to the
present day reaches $\delta_{\rm crit}(z)=1.69[D(z)/D(0)]^{-1}$, where the
function $D(z)$ is the growth factor. In the overdense region surrounding a massive galaxy the growth factor is
modified relative to the background universe because of the enhanced matter
density. Each spherical shell surrounding the central halo behaves as if it belongs to a universe with a
modified value of the density parameter $\Omega_m$ given by its radial
expansion rate and the average matter density at the radius of the
shell. We use the corresponding value of the growth factor for the value of
$\Omega_m$ associated with each shell.

Within the HII region the neutral fraction is close to zero ($x_{\rm
HI}\sim0$).  Beyond the HII region, the clustering of galaxies may still
yield ionisation at a level that is larger than the global average. The
structure of ionisation in the IGM prior to reionisation is thought to be
like that of a sponge, the neutral IGM being punctuated with HII regions
(each with a sharp boundary) of different sizes. An observation at infinite
resolution would be able to resolve all these HII regions. However a real
instrument will observe this sponge-like structure at finite
resolution. Beyond the radius of the HII region we therefore define the
local neutral fraction to be the mass fraction of neutral hydrogen averaged
within a region that is smaller than the size of the main HII region, but
large enough to contain many HII regions around small galaxies. This is the
appropriate measure to consider for observations at finite spatial
resolution (e.g. redshifted 21cm observations).

At a radius $R$ from the central massive galaxy, the neutral hydrogen fraction is
\begin{equation}
\label{neutfrac}
x_{\rm HI}(R) = 1-\int_{M_{\rm min}}^{\infty}dM \frac{4\pi}{3}\frac{D_{\rm M}^3}{\Delta_{\rm
R}^2}\frac{dn\left(R,M_{\rm halo}\right)}{dM}.
\end{equation}  
At large $R$, ${dn\left(R,M_{\rm halo}\right)}/{dM}$ approaches the average
mass-function in the Universe, $\Delta_{\rm R}$ approaches unity, and
equation~(\ref{neutfrac}) reduces to the average neutral fraction in the
IGM (see equation~\ref{FH}). Thus, away from the central HII region the above definition of neutral fraction is equivalent to the mean neutral fraction of the IGM.

\section{Model Parameters and the Reionisation History}

In order to compute the size and profile of an HII region, we need
to specify the star formation efficiency and its dependence on halo mass,
as well as the escape fraction of ionising photons from galaxies. As in
Paper I, we define four models which bracket the expected range of star
formation efficiencies. We then fix the free parameters in these models by
requiring that the resulting reionisation history match observational
constraints.

\subsection{Models for the star formation efficiency}

First we describe the four different models for the star formation
efficiency that are examined in this paper.  We adopt feedback-regulated
prescriptions (Dekel \& Woo~2002; Wyithe \& Loeb~2003) in which the star
formation efficiency in a galaxy depends on its velocity dispersion
$\sigma$, with $f_{\star}=0$ for $\sigma<\sigma_{\rm min}$,
$f_{\star}=f_{\rm \star,crit} \left({\sigma}/{\sigma_{\rm crit}}\right)^2$
for $\sigma_{\rm min}<\sigma<\sigma_{\rm crit}$, and $f_{\star}=f_{\rm
\star,crit}$ for $\sigma>\sigma_{\rm crit}$.  We consider four cases in
this paper: ({\bf A})~$\sigma_{\rm min}=\sigma_{\rm crit}=10~{\rm
km~s^{-1}}$; ({\bf B})~$\sigma_{\rm min}=10~{\rm km~s^{-1}}$, $\sigma_{\rm
crit}=120~{\rm km~s^{-1}}$; ({\bf C})~$\sigma_{\rm min}=\sigma_{\rm
crit}=50~{\rm km~s^{-1}}$; and ({\bf D})~$\sigma_{\rm min}=50~{\rm
km~s^{-1}}$, $\sigma_{\rm crit}=120~{\rm km~s^{-1}}$. The values of
$\sigma_{\rm min}=50~{\rm km~s^{-1}}$, and $\sigma_{\rm min}=10~{\rm
km~s^{-1}}$, correspond to the threshold for infall of gas from a
photo-ionised IGM and the threshold for atomic hydrogen cooling,
respectively.  The value of $\sigma_{\rm crit}=120~{\rm km~s^{-1}}$ is
calibrated empirically based on the local sample of galaxies (Kauffmann et
al.~2003; Dekel \& Woo 2002). Case~{\bf A} is a fiducial model where all
galaxies have the same star formation efficiency, and form down to masses
corresponding to the hydrogen cooling threshold.  This model describes the
situation at high redshift where the IGM was not pre-heated, and where
feedback does not operate in small galaxies as at low redshift because the
dynamical time within galaxies is comparable to the lifetime of massive
stars. Cases~{\bf B-D} introduce modifications due to galaxy formation in a
heated IGM and to stellar feedback.

\subsection{The reionisation history}

As described in Paper I, we may constrain the free parameters in each of
the four models defined above by requiring that hydrogen be reionised by
$z=6$ (Fan et al.~2004), and helium be doubly reionised by $z=3.5$ (Theuns
et al.~2002).  The global ionisation fraction of intergalactic hydrogen is
\begin{equation}
\label{FH}
F_{\rm ion,HII}=\int_{M_{\rm
min}}^{\infty}dM\frac{dn}{dM}\frac{4\pi}{3}D_{\rm M}^3.
\end{equation}
Reionisation histories based on equation~(\ref{FH}) for the different star
formation prescriptions {\bf A-D} which were described in the previous
section are plotted in Figure~\ref{fig1}.  Note that these histories (and
subsequent results) are independent of $N_{\rm reion}$ which is degenerate
with other parameters (e.g. $f_{\rm star}$).

The total optical depth for electron scattering is $\tau_{\rm es}=0.06$ in
case {\bf A} and $\sim 0.05$ in cases {\bf B}--{\bf D}.  All four models
are consistent with the latest measurement of $\tau_{\rm es}= 0.09\pm0.03$
from WMAP (Spergel et al. 2006).  The error bars on the WMAP data do not
allow us to distinguish between the models.

\section{Results}
\begin{figure*}
\includegraphics[width=14cm]{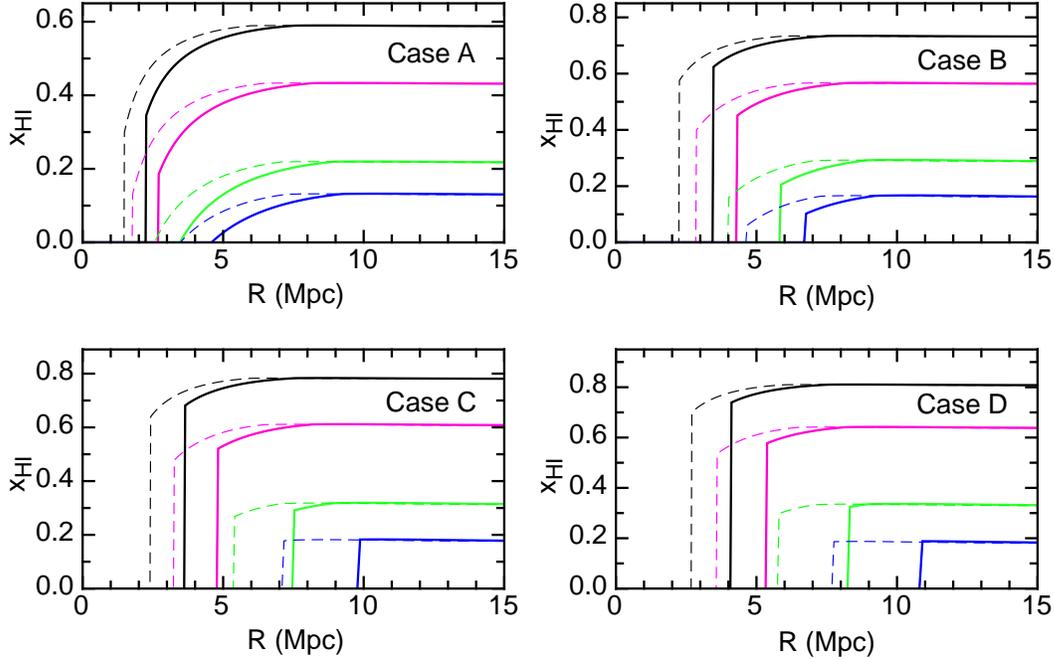} \caption{Neutral fraction as a
  function of radius around a quasar at $z>6$. The four panels show results
  for cases {\bf A}--{\bf D}. In each case the solid and dashed curves show
  central velocity dispersions of $\sigma=350~{\rm km~s^{-1}}$ and
  $\sigma=200~{\rm km~s^{-1}}$ respectively. The four sets of curves from
  top to bottom correspond to redshifts of $z=7.5$, 6.9, 6.3 and 6.1.}
\label{fig2}
\end{figure*}

We now examine our results for the effect of clustered sources on the
neutral fraction outside the HII regions surrounding massive galaxies
(equation~\ref{neutfrac}).
 
\subsection{The neutral fraction outside quasar HII regions}

In Figure~\ref{fig2}, we plot the neutral fraction $x_{\rm HI}$ as a
function of radius $R$ around a central quasar.  The four panels in
Figure~\ref{fig2} refer to cases~{\bf A-D}, and the solid and dashed lines
correspond to velocity dispersions of $350~{\rm km~s^{-1}}$ and $200~{\rm
km~s^{-1}}$ for the central galaxy, respectively. In each case the four
sets of curves correspond to redshifts of $z=7.5$, 6.9, 6.3 and 6.1 (top to
bottom). These calculations of $x_{\rm HI}(R)$ are self-consistent with the
average neutral fraction of the IGM, which is obtained from the
reionisation histories (equation~\ref{FH}). The figure shows that the edges
of HII regions are not expected to be sharp, particularly when the neutral
fraction is of order unity. Indeed, in such cases the neutral fraction
rises gradually beyond $R_{\rm s}$, and reaches the global value only at
radii that are 2--5 times $R_{\rm s}$.

Figure~\ref{fig2} shows that the detailed profile of the neutral fraction
with radius depends on the value of the global neutral fraction. In all
cases the global neutral fraction is reached at a radius
$\sim8$Mpc. However galaxies produce larger HII regions within a more
ionised IGM. As a result, the boundary of the HII region is spread over a
larger range of radii (relative to $R_{\rm s}$) in cases where the global
neutral fraction is higher. The profile is also sensitive to the assumed
case of star formation. In particular, consider star-formation scenarios
where there is only a small contribution to the ionising photon production
from galaxies with $\sigma\la50$km/s. The boundary is less extended in such
cases, particularly where the HII region forms within an IGM that is
already significantly reionised. On the other hand, for large global
neutral fractions the curves corresponding to the four models are
similar. This is simply because the parameters in these models have been
tuned to provide hydrogen reionisation at $z\sim6$. By pinning the models
to the same redshift of reionisation, we draw the robust conclusion that
HII regions around high redshift quasars within a substantially neutral IGM
should not have sharp edges, but rather the neutral fraction in the IGM
will gradually reach the global value.

We note that the effect on the ionisation fraction is significant relative
to the background level of ionisation on scales as large as $\sim$30
comoving Mpc. Although the overdensity on this scale is mild, $\sim10\%$,
it nevertheless biases the abundances of galaxies, particularly those on
the exponential tail of the mass function, to form earlier (Barkana \& Loeb
2004). This early formation produces an ionisation level that is
significantly larger than that of the background.  The magnitude of the
effect we calculate agrees with expectations for the clustering of galaxies
(Scannapieco \& Barkana~2002). For example, given a halo mass of $\sim
10^9M_\odot$ (that is a factor $\sim 10^3$ smaller than a central galaxy
with $\sigma=350$km/s at $z=7.5$), we find the cross-correlation length
with the central halo to be $\sim1.3$ physical Mpc.  This length represents
density enhancements of order unity and is naturally smaller than the
$\sim3$--$5$Mpc where enhanced ionisation is seen outside the HII
region. However we find a 50\% enhancement in the density of halos at
$\sim2$Mpc (physical scale) relative to a random distribution, and a 10\%
enhancement at $\sim4$Mpc.  For a central galaxy with $\sigma=200$km/s, we
find a correlation length of $\sim0.8$Mpc, with an enhancement of 50\% at a
distance of $\sim1.3$Mpc and of 10\% at $\sim3$Mpc.  These enhancements can
naturally account for the 10--30\% increase in the ionisation level at
these radii.

\subsection{The neutral fraction outside galaxy HII regions}

\begin{figure*}
\includegraphics[width=14cm]{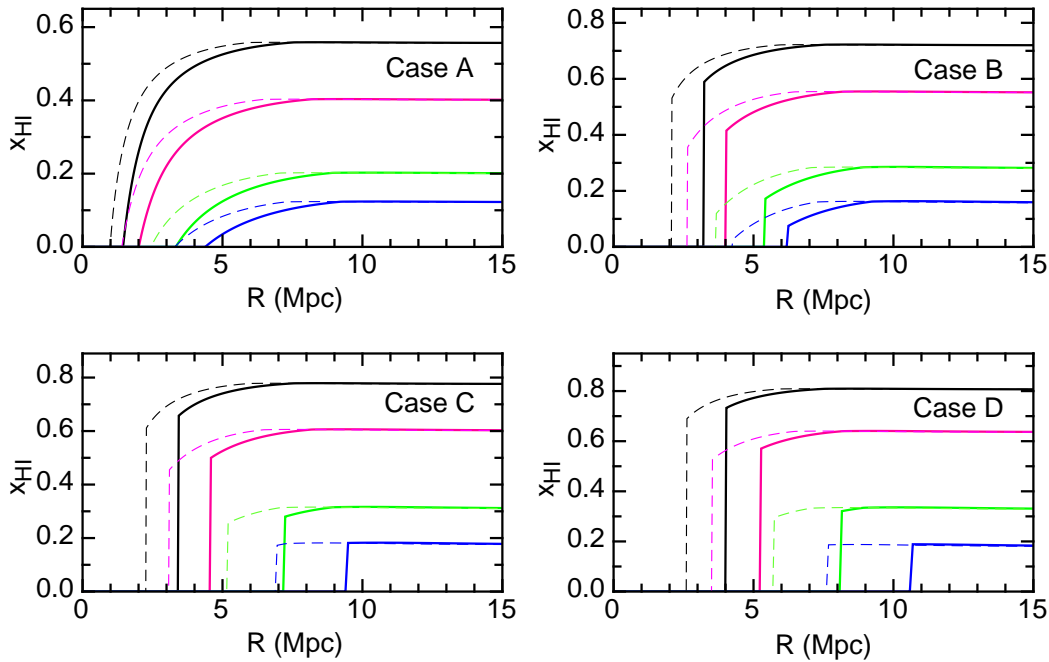} \caption{Neutral fraction as a
  function of radius around a massive galaxy at $z>6$ in the absence of
  quasar activity. The four panels show results for cases {\bf A}--{\bf
  D}. In each case the solid and dashed curves show central velocity
  dispersions of $\sigma=350~{\rm km~s^{-1}}$ and $\sigma=200~{\rm
  km~s^{-1}}$ respectively. The four sets of curves from top to bottom
  correspond to redshifts of $z=7.5$, 6.9, 6.3 and 6.1. }
\label{fig3} 
\end{figure*}

In Paper I we demonstrated that quasars do not generally dominate the
contribution to ionisation within an HII region around clustered galaxies.
Therefore even in the absence of quasar activity we would expect a large
HII region surrounding a massive galaxy. Clustering of galaxies around a
central massive galaxy should also lead to a gradual boundary of the HII
region as is the case for a central quasar.

We therefore repeat the calculations of the previous section in the case
where quasar activity is ignored (both in the central galaxy, and
surrounding galaxies), by replacing $D_{\rm halo}^3=D_{\rm halo,q}^3+D_{\rm
halo,\star}^3$ and $D_{\rm M}^3=D_{\rm M,q}^3+D_{\rm M,\star}^3$, with
$D_{\rm halo}=D_{\rm halo,\star}$ and $D_{\rm M}=D_{\rm M,\star}$ in
equations~(\ref{RS})-(\ref{FH}). The results are shown in
Figure~\ref{fig3}. We find profiles that are qualitatively consistent with
those shown in Figure~\ref{fig2}. The quantitative difference is that HII
regions around massive galaxies are smaller in the absence of quasars. As
before, the neutral fraction rises gradually beyond $R_{\rm s}$, and
reaches the global value only at radii that are 2--5 times $R_{\rm s}$.

\section{Summary}

Our calculations suggest that observations of cosmological HII regions will
not see sharp boundaries, but rather a gradual rise in the average local
neutral fraction from the radius of the HII region out to a distance
$2$--$5$ times greater. This gradual rise in neutral fraction is produced
by the clustering of sources around the massive systems at the centers of
cosmological HII regions. The gradual rise is seen in our models for HII
regions surrounding high redshift quasars, as well as massive galaxies in
the absence of quasar activity. The gradual nature of the HII region
boundary is most prominent in star formation scenarios that do not include strong feedback in low mass galaxies.

Our {\it Letter} only considered the softening of the HII region boundary
by galaxy clustering. Additional softening of the boundary could be
produced near quasars that emit significant energy in X-ray photons, whose
mean-free-path is much longer than UV photons. For an unobscured quasar the
spectrum averaged mean-free-path is much smaller than the size of the HII
region. As a result, the high energy photons smooth the edge of the HII
region over a length much smaller than its radius.  Along directions where
the UV emission is obscured but the X-rays are not, the shape of the HII
boundary might be dominated by the quasar spectrum and not the clustering
of galaxies as discussed here.

We have assumed spherical symmetry in our model. However, in reality
massive galaxies and quasars will lie at the intersections of filaments
along which mass is funneled into them. The shape of the HII region and the
level to which the boundary is gradual, will vary along different
directions. Numerical simulations are needed to explore this anisotropy.
Nevertheless, the first observations of HII regions at redshifted 21cm
wavelengths will be made at a resolution comparable to the size of the HII
region in both transverse and line-of-sight directions (Wyithe, Barnes \&
Loeb~2005). This coarse resolution will tend to smooth the neutral
distribution in spherical shells as assumed in our analysis.

The first generation of low-frequency arrays will likely detect HII regions
via matched filter techniques. In the case of HII regions also
observed through Ly$\alpha$ absorption, the clustering of sources results
in a "characteristic size" of an HII region (as measured assuming a sharp
boundary) that is 1.5 to 2 times greater than the value measured via
Ly$\alpha$ absorption. Searches for HII regions using redshifted 21cm
observations around known quasars should therefore use a filter that
accounts for the gradual profile of the HII region boundary.

The detection of the distribution of HII region sizes, together with the
power-spectrum of 21cm emission will provide a detailed picture of the
topology of the reionisation epoch (Furlanetto et al.~2004). We have
demonstrated that there is a need to define the size of an HII region in a
resolution dependent way. In particular, finite resolution observations
(for example using low-frequency radio telescopes to detect redshifted 21cm
radiation) will not detect sharp boundaries around large HII
regions. Rather, confusion of many small, clustered and unresolved HII
regions, will blur the observed boundary of the HII regions around massive
galaxies and quasars, and potentially increase their inferred HII volumes
by an order of magnitude.

{\bf Acknowledgments} The research was supported by the Australian Research
Council (JSBW) and NASA grants NAG 5-1329 and NNG05GH54G (AL).

\newcommand{\noopsort}[1]{}

\label{lastpage}
\end{document}